\documentstyle[aps,twocolumn,amsmath,amssymb]{revtex}
\newcommand{\be}{\begin{equation}}
\newcommand{\ee}{\end{equation}}
\newcommand{\bea}{\begin{eqnarray}}
\newcommand{\eea}{\end{eqnarray}}
\newcommand{\ket}[1]{\left| #1\right \rangle}
\newcommand{\bra}[1]{\left \langle #1\right|}
\newcommand{\twoket}[2]{\left| #1\right \rangle
                        \otimes
                        \left| #2\right \rangle }
\newcommand{\twobra}[2]{\left \langle #1\right|
                        \otimes
                        \left \langle #2\right|}
\newcommand{\scal}[2]{\langle #1| #2 \rangle}
\newcommand{\av}[1]{\langle #1 \rangle}
\newcommand{\aval}[1]{\langle #1 
        \rangle_{\{\al{},\ald{},\fs{},\ms{},\ns{}\}}}
\newcommand{\avnull}[1]{\langle #1
        \rangle_{\{0,0,\fs{},\ms{},\ns{}\}}}
\newcommand{\trace}[1]{\textrm{Tr}\left\{ #1 \right\}}
\newcommand{\comut}[2]{[#1,\,#2\,]}

\newcommand{\al}[1]{\alpha_{#1}}
\newcommand{\ald}[1]{\alpha_{#1}^\ast}

\newcommand{\mf}{\ket{\al{}}}
\newcommand{\mfd}{\bra{\al{}}}

\newcommand{\aop}[1]{{\hat{a}_{#1}}}
\newcommand{\aopd}[1]{{\hat{a}_{#1}^\dag}}

\newcommand{\gamopi }{\hat{\gamma}_{i}}

\newcommand{\gamopj}{\hat{\gamma}_{j}}

\newcommand{\gamopset}[1]{\{  \hat{\gamma}_{#1}|\,#1 \in {\cal I}\}}

\newcommand{\gami }{\gamma_{i}}

\newcommand{\gamj}{\gamma_{j}}

\newcommand{\Upsq}[1]{{\Upsilon}^{#1}_{ \{\fs{},\ms{},\ns{}\}}}
\newcommand{\Lamsq}[1]{{\Lambda}^{#1}_{\{\fs{},\ms{},\ns{}\}}}

\newcommand{\Gcoll}[1]{{\Gamma_{#1}}}

\newcommand{\bgami }{\overline{\gamma}_{i}}

\newcommand{\sigam}[1]{{\mathbf{\sigma}}_{\{\gamma #1\}}}

\newcommand{\brho}{{\mathbf{\rho}}}

\newcommand{\signal}{{\mathbf{\sigma}}_{
        \{\al{},\ald{},\fs{},\ms{},\ns{}\}}^{(0)}}
\newcommand{\sign}[1]{{\mathbf{\sigma}}_{\{\gamma #1\}}^{(0)}}

\newcommand{\signSchroedinger}[2]{
        \sigma_{ \{\overline{\gamma}(#1;\{\gamma #2\})\}}^{(0)}}

\newcommand{\sigo}[1]{{\mathbf{\sigma}}_{\{\gamma #1\}}^{(1)}}

\newcommand{\Undag}[1]{\left. {{\widehat{U}}_{\{\gamma\}}^{(0)}}\!\!\right. 
^{\dagger}\!\!(#1)}
\newcommand{\Un}[1]{\widehat{U}_{\{\gamma\}}^{(0)}(#1)}

\newcommand{\bx}{{\mathbf{x}}}
\newcommand{\bp}{{\mathbf{p}}}
\newcommand{\by}{{\mathbf{y}}}

\newcommand{\hng}[1]{\widehat{H}^{(0)}_{\{\gamma #1\}}}

\newcommand{\hog}[1]{\widehat{H}^{(1)}_{\{\gamma #1\}}}
\newcommand{\hogSchroedinger}[2]{
        \widehat{H}_{ \{\overline{\gamma}(#1;\{\gamma #2\})\}}^{(1)}}
\newcommand{\Qog}[1]{\widehat{Q}^{(1)}_{\{\gamma #1\}}}

\newcommand{\hn}{\widehat{H}^{(0)}}
\newcommand{\ho}{\widehat{H}^{(1)}}
\newcommand{\astruc}[1]{{{\mathcal{A}}_{\{{#1}\}}}}
\newcommand{\psiop}[1]{\hat{a}_{#1}}
\newcommand{\psiopd}[1]{\hat{a}_{#1}^\dag}
\newcommand{\kpsiop}[1]{\hat{\psi}_{#1}}

\newcommand{\ppr}{{\prime \prime}
}
\newcommand{\fs}[1]{\tilde{f}_{#1}}
\newcommand{\fsop}[1]{{\,\,\hat{\!\!\tilde{f}}_{#1}}}

\newcommand{\fspo}[1]{(1+\tilde{f})_{#1}}

\newcommand{\ms}[1]{\widetilde{m}_{#1}}
\newcommand{\msop}[1]{{\hat{\widetilde{m}}}_{#1}}
\newcommand{\ns}[1]{\widetilde{n}_{#1}}
\newcommand{\nsop}[1]{\hat{\widetilde{n}}_{#1}}

\newcommand{\fc}[1]{f^{(c)}_{#1}}
\newcommand{\mc}[1]{m^{(c)}_{#1}}
\newcommand{\nc}[1]{n^{(c)}_{#1}}
\newcommand{\ft}[1]{f_{#1}}
\newcommand{\mt}[1]{m_{#1}}
\newcommand{\nt}[1]{n_{#1}}

\newcommand{\phiunpr}{\phi^{1 2 3 4}}
\newcommand{\phipr}{\phi^{1\,2^\prime 3^\prime 4^\prime}}
\newcommand{\phippr}[1]{\phi^{1^\ppr 2^\ppr 3^\ppr 4^\ppr}_{#1}}

\newcommand{\dd}[1]{{\partial_{ #1}}} 

\newcommand{\Uc}{U_{\fc{}}}
\newcommand{\Usq}{U_{\fs{}}}

\newcommand{\Vsq}{V_{\ms{}}}
\newcommand{\Vm}{V_{(\mc{}+\ms{})}}
\newcommand{\Ltwo}[1]{L_{\{\al{},\ald{},\fs{},\ms{},\ns{}\}}^{(2)}[#1]}

\begin{document}
\draft
\wideabs{ 
  \title{Quantum Kinetic Theory for a Condensed Bosonic Gas}
  \author{R. Walser, J. Williams, J. Cooper and M. Holland}
  \address{JILA, National Institute for Standards and Technology and
    University of Colorado, Boulder, CO 80309-0440}
  \date{February 16., 1999}
  \maketitle{}

  \begin{abstract}
    We present a kinetic theory for Bose-Einstein condensation of a
    weakly interacting atomic gas in a trap. Starting from first
    principles, we establish a Markovian kinetic description for
    the evolution towards equilibrium.  In particular, we obtain a set
    of self-consistent master equations for mean fields, normal
    densities, and anomalous fluctuations. These kinetic equations
    generalize the Gross-Pitaevskii mean-field equations, and merge
    them consistently with a quantum-Boltzmann equation approach.

  \end{abstract}
 }


\narrowtext

\section{Introduction}
With the experimental realization of Bose-Einstein condensation with
neutral atomic gases \cite{cornellsci,hulet895,ketterle1195}, it has
become possible to observe fundamental properties of quantum
statistics directly. While there are numerous well-established fields
of low-temperature quantum physics that deal with many-particle
systems, most of the salient features related to the paradigm of
indistinguishablility are masked by strong interactions, or contact
with thermalizing reservoirs.  The newly gained ability to isolate the
condensate fraction from its environment to a high degree, as well as
the tight spatial confinement of the macroscopic quantum field, has
opened up ways to selectively manipulate the mean-field, and study its
temporal evolution towards equilibrium almost "in vivo".

The motion of trapped atoms in a dilute gas consists of free
oscillations within the external potential that are interrupted by
short binary collision events. Conventionally, that interaction
strength is measured by the range of the repulsive two-particle
potential, i.e. the scattering length $a_{{\text s}}$.  The duration
of a collision event $\tau_0$ is given by the time a particle of
average velocity $v$ spends in the interaction region, i.e.,
$\tau_0=a_{{\text s}}/v$. On the other hand, the inverse collision
rate $\tau_{\text{c}}=1/(n\,a_{{\text s}}^2\,v)$ is an estimate of the
time between successive collisions where $n$ denotes a particle
density.  As we are interested in low kinetic energies and the weak
interaction limit, one finds a characteristic separation of time
scales, i.e.,  \bea
\label{gaseousparam}
&&\tau_{{\text c}} \gg \tau_0, \quad \mbox{or}\quad n a_{{\text s}}^3
\ll 1.  \eea 
It is this separation of time-scales that gives raise to
a kinetic stage of evolution, preceding any equilibrium situation.
For example, such a kinetic stage is absent in strongly interacting
systems where a local equilibrium is established immediately
($\tau_{{\text c}}\approx \tau_0$, hydro-dynamic stage).  Any
individual collision event creates a quantum mechanical entanglement
between collision partners.  However, due to the long separation
between successive collisions and the presence of intermediate weak
fluctuations these temporal correlations decay rapidly (Markov
approximation).  

Based on this microscopic picture of the weakly interacting bosonic gas, 
we derive a generalized kinetic theory for a coarse-grained Markovian 
many-particle density operator, as discussed by
{\em A.~I. Akhiezer and S.~V. Peletminskii},\cite{peletminskii}. 
The quint-essential assumption behind this approximation is a coarse-grained
density operator that depends only on a few selected variables. 
These quantities will serve as master variables and 
determine the system's evolution on a coarse-grained time scale. 
From a  perturbative expansion of this density operator,
we obtain kinetic equations that describe the temporal evolution of the 
expectation value of any (single-time) observable.
As a specific application  of this formulation, we then 
assume that the condensed gas can be described essentially 
by the dynamic evolution of mean fields, normal- and anomalous fluctuations. 
In particular, we obtain kinetic equations that
comprise the Gross-Pitaevskii equation and the quantum-Boltzmann 
equation as special instances.
These kinetic equations are formulated  basis-independently  and 
consider all second-order processes which give raise to
collisional energy shifts and damping rates. 
It is also noteworthy that the present
theory can be applied readily to multi-component bosonic gases, provided
the two-particle scattering matrix is interpreted accordingly.   

However, considering the rapid development of this field of physics,
it is also of great importance to compare the predictions of this and
other kinetic theories
\cite{walraven96,kagan95,proukakis96,holland196,zolleri,zollerii,williams597,gardiner997,zolleriii,zolleriv,smerzi798}
(mean-field equations and kinetic theories) with the results obtained
from finite-temperature calculations based on the Hartree-Fock-Bogoliubov
 formulation
\cite{javanainen1196,griffin496,griffin397,stringari1197,dodd198,fetter298,giorgini498,dodd998,shlyapnikov1098,laloe98}
(collective excitation frequencies and damping rates), direct
approaches to solve the many-particle Schr{\"o}dinger equation
\cite{esry97,greene798} (configuration interaction, hyper-spherical
coordinates), renormalization group techniques \cite{stoof1296}, and,
above all, to gauge them against physical reality
\cite{cornell796,ketterle896,cornell297}.  Further references on the
extensive literature can be found in recent review articles and
literature compilations
\cite{parkinsreview,stringarireview,shlyapnikovreview}.
  
The present article is organized as follows: In Sec.~\ref{sec2a}, we
introduce a coarse-grained statistical density operator and derive an
integral equation for it in Sec.~\ref{sec2b}. Assuming a weak
two-particle collision rate, permits the development of a perturbation
theory, which is described in Sec.~\ref{sec2c}.  From an explicit series 
expansion of the
coarse-grained density operator, we obtain in Sec.~\ref{sec2d} a set
of kinetic equations characteristic of a condensed bosonic gas.  In
Sec.~\ref{sec3a}, we introduce the Hamilton operator that governs the
kinetic evolution of a weakly interacting, repulsive gas.  A set of
relevant operators is introduced in Sec.~\ref{sec3b}, i.e.,
macroscopic mean-fields, normal fluctuations, as well as anomalous
averages.  By applying the general kinetic master equations to this
specific set of relevant operators, we obtain self-consistent kinetic
master-equations and give a detailed discussion in Sec.~\ref{sec3c}.
Finally, in Sec.~\ref{sec4}, we discuss the prospects and possible 
implications for a numerical simulation of a 
fully quantum mechanical self-consistent solution. 

\section{A coarse-grained Markovian density operator}
\label{sec2}
The quantum mechanical state of a many-particle system contains an
overwhelming amount of data and it is the objective of quantum
statistical mechanics to extract the relevant information from a data
subset as small as possible. Such a minimal description then leads
to a coarse-grained picture of the behavior of the system.  The
effects of coarse-graining on the dynamical evolution are well known
from the quantum mechanics of open systems \cite{gardiner,carmichael}.
By discarding some observable information from the unitary evolution
of the complete system, one breaks the time-reversal symmetry, thus
introducing irreversibility.  However, a judicious partitioning of the
many, interacting degrees of freedom of a large system into a small
relevant subsystem and a weakly coupled, complementary reservoir
leads to a tractable description of the subsystem's evolution towards
equilibrium.  This equilibrium state is determined by the properties
of the reservoir.

In this section, we will pursue these general ideas and define a
coarse-grained statistical density operator that depends functionally
on only a few fundamental variables, and thus gives a raw picture of
the ``true'' state of the many-particle system.  Based on the
definition for the coarse-grained statistical density operator, we
then derive an integral equation that determines the functional form
of this operator. In the limit of weakly interacting, dilute quantum
gases where strong collisional interaction events are well separated
in time, it is possible to solve this equation perturbatively and
establish a hierarchy in terms of an expansion parameter proportional
to this interaction strength.  Once a perturbative expression for the
functional form of the coarse-grained statistical operator is
determined, we then use it to study the motion of expectation values
of general observables.  These equations of motion establish a closed,
self-consistent set of kinetic equations for the following restricted
set of fundamental variables: the mean field, the normal
single-particle density, and the anomalous fluctuations.  By this
method, we generalize the Gross-Pitaevskii mean-field equation
and merge it consistently with an approach that leads to the
quantum-Boltzmann equations for both normal densities and
anomalous fluctuations.

In many ways our approach is reminiscent of the classical 
Bogoliubov-Born-Green-Kirkwood-Yvon (BBYGK) method \cite{zubarev1},
for in a similar manner, higher order correlations can also be
characterized by a restricted set of variables. Specifically, in
classical Markovian systems higher order correlation functions
can be expressed in terms of single-particle densities.

\subsection{Fundamental assumptions of statistical mechanics}
\label{sec2a}
The derivation of a coarse-grained density operator relies on the
basic assumption of statistical mechanics that any non-equilibrium
statistical correlation eventually decays
\cite{peletminskii,zubarev1,chapman,huang,zubarev2}.  For example, in
classical kinetics, this time is typically of the order of the duration of
a collision.
  
In the following discussion, we introduce three distinct many particle
density operators.  The first density operator, $\brho(t)$, describes
the complete state of the many-particle system.  Starting from an
initial value $\brho(0)$, the system evolves unitarily in time
according to a time independent Hamilton operator $\widehat{H}$.  In
analogy with the celebrated Chapman-Enskog \cite{chapman,huang}
procedure of classical statistical mechanics which was later
introduced to quantum statistics by Bogoliubov
\cite{peletminskii}, we assume that all characteristic features of the
quantum statistical state $\brho(t)$ can be inferred from expectation
values of a certain, restricted set of operators $\gamopset{i}$, where
the index $i$ enumerates linearly independent operators $\gamopi$ from
a (possibly infinite) index set ${\cal I}$.

Thus, we introduce a second coarse-grained, Markovian density
operator $\sigma_{\{ \gami(t)|\,i \in {\cal I} \} }$ that
approximates the complete density operator $\brho(t)$ \bea
\label{defsig}
\sigma_{ \{ \gami(t)|\,i\in {\cal I} \} } &\approx&
\brho(t)=e^{-i\widehat{H}t}\, \brho(0)\,e^{i \widehat{H}t}.  \eea 
The
set of expectations values $\{\gami(t)| \,i \in {\cal I}\}$ that
parameterize the coarse-grained density operator are found by a
quantum average $\av{\ldots}$ over all states of the many-particle
configuration space \bea
\label{quantumav}
\gami(t)&=& \av{\gamopi}=\trace{\gamopi \, \brho(t) }= \trace{\gamopi
  \, \sigma_{ \{ \gamma_{j}(t)|\,j \in {\cal I} \} }}.  \eea While the
coarse-graining assumption, i.e., the restriction to a few selected
variables, is a statistical statement, the Markovian postulate
concerns the separation of the time scales that govern the processes
of equilibration and the decorrelation of fluctuations.  Within this
limit, the temporal evolution of this coarse-grained statistical
density operator $ \sigma_{ \{ \gami(t)|\,i \in {\cal I} \} } $ is
solely governed by the motion of the set of expectation values
$\{\gami(t)|\,i \in {\cal I}\}$.  Any time dependence of the matrix
elements of $\sigam{(t)}$ can be attributed to the evolution of this
restricted set of operators $\{ \gamopi \}$, thus no {\em relevant}
intrinsic time dependence is unaccounted for \footnote{In the absence
  of ambiguities, we simplify the notation of the set of operators or
  their corresponding expectation values by dropping the index set
  ${\cal I}$, the indexing label $i$, or even the time argument $t$,
  completely.}.  Although, the set of operators is unspecified so far,
it can include operators such as unity (${\bf 1}$), number
($\hat{N}$), linear momentum ($\hat{P}$), angular momentum
($\hat{L}$), and energy ($\widehat{H}$).  The larger the set of
operators, the better the quality of any subsequent approximation.

The third relevant many-particle density operator is $\sign{}$.  It
serves as a reference distribution and describes a relaxed (but
non-equilibrium) state of the gas between consecutive collision
events.  We define it to yield the same expectation values as
$\sigam{}$ on the restricted set of operators $\{\gamopi\}$.  Thus,
this reference distribution is given by \bea
\label{refdist}
\sign{}&=&\exp{(\gamopi\,\Upsilon^i_{\{\gamma\}})},\eea where repeated
indices imply a summation over operators $\gamopi$ and their conjugate
thermo-dynamic coordinates $\Upsilon^i_{\{\gamma\}}$.  These
coordinates are defined implicitly by \bea
\label{coord2}
\gami&=&\trace{\gamopi \, \sigam{} }= \trace{\gamopi\, \sign{}}.  \eea
It is of interest to note that the reference state, as given by the
exponentiated form in Eq.~(\ref{refdist}), is still of the most
general form permitted by the principles of quantum mechanics, as long
as the set of operators $\{\gamopi\}$ is complete.

This ansatz for a non-equilibrium coarse-grained statistical operator
$\sigam{}$ and a "self-adjusting" reference distribution $\sign{}$
is also used in various contexts of equilibrium thermo-dynamics.  For
example, from a finite order truncation of a quantum virial expansion
\cite{peletminskii}, one obtains a coarse-grained approximation
$\sigma_{\{\Omega,\mu,\beta\}}$ of the grand-canonical statistical
density operator $\brho=\exp{(\Omega-\mu\hat{N}-\beta \widehat{H})}$.
The starting point of the iteration procedure is a reference
distribution $\sigma^{(0)}_{\{\Omega,\mu,\beta\}}$ that matches the
solution as closely as possible. In here, the conjugate thermo-dynamic
coordinates $\{\Omega,\mu,\beta\}$ correspond to the observables:
unity, number and total energy.

\subsection{Derivation of an integral equation}
\label{sec2b}
To find the functional form of the postulated coarse-grained Markovian
 density operator $\sigam{(t)}$, we use a
"boot-strapping" method.  First, by assuming a given, yet unknown
solution of Liouville's equation \bea
\label{Liouville}
\frac{d}{dt} \sigam{(t)}=-i\,\comut{\widehat{H}}{\sigam{(t)}}, \eea we
can determine the dynamical evolution of the expectation values
$\gami(t)=\text{Tr}\{\gamopi \,\sigam{(t)}\}$ from \bea
\label{dynamicmean}
\frac{d}{dt}\gami(t)&=& i \,\trace{\comut{\widehat{H}}{\gamopi}
  \,\sigam{(t)}}.  \eea Second, to derive the functional form of
$\sigam{(t)}$, we now use Eq.~(\ref{Liouville}),
Eq.~(\ref{dynamicmean}), and the fact that there is no explicit time
dependence in $\sigam{(t)}$. Thus, by partial differentiation one
obtains the following equation \bea
\label{Liouvilleneu}
\trace{\comut{\widehat{H}}{\gamopi}\,\sigam{}} \, \dd{\gami}\sigam{} 
&=& -\comut{\widehat{H}}{\sigam{}}, \eea where again, we
have adopted the convention that repeated indices imply summation,
unless stated otherwise.  Moreover, by dropping the explicit time
dependence of the reference point in phase-space $\{\gami{(t)}\}$, one
can consider this as a partial-differential equation with independent
variables $\{\gami{}\}$.

The total Hamilton operator $\widehat{H}$ that governs the evolution
of a weakly interacting, dilute gas permits a partitioning of the
energy into a free part $\hn$ and a presumably weak interaction $\ho$
\bea \widehat{H}&=&\hn+\ho.  \eea One could use the bare,
single-particle energy $\hn$ that determines the free kinetic
evolution of the gas as a starting point of a series expansion of the
coarse-grained density operator in terms of the interaction strength.
However, it is well known that the mean-field interaction will
significantly affect the single-particle energies.  Anticipating this,
we will shift the expansion point $\hn$ by an as yet undetermined
single-particle energy $\Qog{}$ to a dressed energy \bea
\label{hn}
\hng{}&=&\hn+\Qog{}.  \eea To conserve energy, we have to reduce the
interaction energy by an equal amount \bea
\label{ho}
\hog{}&=&\ho-\Qog{}.  \eea The shifted interaction energy $\hog{}$
represents the fluctuations about the mean-field energy and will be
considered as ``weak'', or in other words, that strong fluctuations
are well separated in time.  This procedure is analogous to the first
order energy shift found with ordinary Schr\"odinger-Rayleigh
perturbation theory.  The explicit form of this single-particle
renormalization energy $\Qog{}$ which is of the same order as the
interaction energy, will be determined in the course of this
calculation.

By expanding Eq.~(\ref{Liouvilleneu}) around the dressed
single-particle energy $\hng{}$, one obtains 
\bea
\label{inhompde}
\trace{\comut{\hng{}}{\gamopi}\,\sigam{}}
\, \dd{\gami}\sigam{} 
+\comut{\hng{}}{\sigam{}}
=F_{\{\gamma\}}^{(1)}, 
\eea 
where
$F_{\{\gamma\}}^{(1)}$ is introduced for convenience to hold the
remaining first order contributions 
\bea 
F_{\{\gamma\}}^{(1)}&=&
-\comut{\hog{}}{\sigam{}} 
-\trace{\comut{\hog{}}{\gamopi}\,\sigam{}}
\, \dd{\gami}\sigam{}.  
\eea 
There are two
interesting points in considering Eq.~(\ref{inhompde}).  First, there
is not one unique solution $\sigam{}$, but solution manifolds that are
labeled by the constants of motion.  Thus, any particular solution has
to be augmented with appropriate physical boundary conditions.
Second, the structure of the commutator of the single-particle
Hamiltonian $\hng{}$ with the set of operators $\{\gamopi\}$ is an
intrinsic property of the system.  It is determined both by the
particular physical configuration and the set of operators. If the set
of operators is chosen appropriately, the commutator forms an algebra
with structure constants $\astruc{\gamma}$ defined by 
\bea
        \label{struct}
        \comut{\hng{}}{\gamopi}&=&{{\astruc{\gamma}}_{i}}^j \,\gamopj.
        \eea 
If we substitute this algebraic closure relation into
 Eq.~(\ref{inhompde}), one obtains 
\bea
\label{inhompde2}
({\astruc{\gamma}}^{ij}\,\gamj)
\,\dd{\gami}\sigam{} 
+\comut{\hng{}}{\sigam{}}=F_{\{\gamma\}}^{(1)} .  \eea

By the method of characteristics, one can transform this first order,
inhomogeneous partial-differential equation into an equivalent
integral equation. Although formally equivalent, the later method is
advantageous as it leads naturally to a series expansion of the
coarse-grained density operator by iteration (compare Dyson series 
\cite{greiner}).  The characteristic trajectories are
parameterized curves $\{\bgami(\tau;\{\gamma\})\}$ along which the
boundary value of the coarse-grained density operator is propagating
in phase-space according to Eq.~(\ref{inhompde2}).  They are defined
by \bea
\label{characteristics}
\frac{d}{d\tau}\bgami(\tau;\{\gamma\})&=&
i\,{{\astruc{\overline{\gamma}(\tau;\{\gamma\})}}_{i}} ^{j}
\,\overline{\gamma}_{j}(\tau;\{\gamma\}),\\
\bgami(\tau=0;\{\gamma\})&=&\gamma_{i}.  \eea and the corresponding
boundary conditions.  Formally, the solution of the differential
equation defines a map $K_{\{\gamma\}}(\tau)$ from regions in
phase-space connected by the characteristics, i.e.,  \bea
\bgami(\tau;\{\gamma\})&=& {K_{\{\gamma\}}(\tau)_{i}}^{j}\,\gamma_{j}.
\eea This map is obtained from the equation defining the
characteristics Eq.~({\ref{characteristics}), i.e.,  \bea
\label{map}
\frac{d}{d\tau}{K_{\{\gamma\}}(\tau)_{i}}^{j}&=&
i\,{{\astruc{\overline{\gamma}(\tau;\{\gamma\})}}_{i}}
^{l}\,{K_{\{\gamma\}}(\tau)_{l}}^{j},\\
K_{\{\gamma\}}(\tau=0)&=&{\bf 1}.  \eea As the coarse-grained density
operator is transported along the characteristics from its boundary
value to its final value, it also evolves freely according to
Eq.~(\ref{inhompde2}).  To account for this evolution, we also define
a unitary propagator $\Un{\tau}$ by \bea
        \label{propagator}
        \frac{d}{d\tau} \Un{\tau}&=& -i\,
        \widehat{H}_{\left\{\overline{\gamma}(\tau;\{\gamma\})
          \right\}}^{(0)}\,
        \Un{\tau},\\
        \Un{\tau=0}&=&{\bf 1}.  \eea The solution for the
        inhomogeneous partial-differential equation
        Eq.~(\ref{inhompde2}) is now obtained from an
        interaction-picture representation of the density operator
        $\sigam{}(\tau)$ that is defined by \bea
\label{aux}
\sigam{}(\tau)&=&\Undag{\tau} \;
\sigma_{\{\overline{\gamma}(\tau;\{\gamma\})\}} \, \Un{\tau}.  \eea At
$\tau=0$, it coincides with the Schr\"odinger-picture value \bea
\label{boundarypresent}
\sigam{}(\tau=0)=\sigma_{\{\gamma\}}, \eea and, if our reference
distribution, as defined in Eq.~(\ref{refdist}) matches the input
state in the remote past asymptotically, i.e.,  \bea \lim_{\tau
  \rightarrow -\infty}
\sigma_{\{\overline{\gamma}(\tau;\{\gamma\})\}}\approx \lim_{\tau
  \rightarrow -\infty} \signSchroedinger{\tau}{}, \eea then we find
for the initial condition that \bea
        \label{boundarypast}
        \lim_{\tau\rightarrow -\infty} \sigam{}(\tau)=\sign{}.  \eea
        Finally, by differentiating the interaction-picture
        representation Eq.~(\ref{aux}), the use of
        Eq.~(\ref{inhompde2}) and a subsequent integration subject to
        the boundary conditions given in
        Eqs.~(\ref{boundarypresent},\ref{boundarypast}), the integral
        equation for the coarse-grained density operator is obtained:
        \bea
        \label{integral}
        \sigam{}&=&\sign{} -i\,\int_{-\infty}^0 d\tau\, e^{\eta
          \tau}\, \Undag{\tau}\, \left(
          \comut{\hog{}}{\sigam{}}+\right.\nonumber\\
        &+&\left.
\trace{
            \comut{\hog{}}{\gamopi}\, \sigam{}}
\,\dd{\gami}\sigam{}
        \right)_{|\overline{\gamma}(\tau;\{\gamma\})}
        \Un{\tau}. \nonumber\\
        \eea This expression links the interacting coarse-grained
        density operator to its non-interacting value by evaluating
        the integrand along its collision history. A regularizing
        function ($\eta\rightarrow 0_+$) has been introduced and
        suppresses higher order correlations, built up in individual
        collisional events.

\subsection{The limit of weakly interacting, dilute gases}
\label{sec2c}
In the case of a weakly interacting system, we can seek the solution
to the integral equation Eq.~(\ref{integral}) in form a power series
of the density operator. The expansion parameter is given by the
interaction strength \bea
\label{series}
\sigam{}&=&\sum_{l=0}^{\infty}\sigam{}^{(l)}.  \eea 
If this series is
inserted into the integral equation for the coarse-grained density
operator, then one finds for the first order correction 
\bea
\label{firstorder}
\sigo{}&=& -i\,\int_{-\infty}^0 d\tau\, e^{\eta \tau}\, \Undag{\tau}\,
\left( \comut{\hog{}}{\sign{}}+
\right.\\
&+&\left.  
\trace{ \comut{\hog{}}{\gamopi}\,\sign{}} 
\,\dd{\gami}\sign{}
\right)_{|\overline{\gamma}(\tau;\{\gamma\})}
\Un{\tau}.\nonumber \eea We will show that the first two terms of the
series Eq.~(\ref{series}) are sufficient to determine the kinetic
equations for the expectation values of the set of operators
$\{\gamopi\}$ including second order (collisional) contributions.
        
It is important to note that the integrand has to be evaluated along
the trajectories to the past. However, it can be further simplified by
using again the interaction-picture representation, as derived from
Eqs.~(\ref{struct},\ref{map},\ref{propagator}).  First, one notices
that the interaction-picture representation of the reference
distribution is constant, if evaluated along the characteristic
trajectories.  This is implied in its definition.  \bea \sign{}&=&
\Undag{\tau}\,\signSchroedinger{\tau}{} \,\Un{\tau}.  \eea Second, one
finds that in the interaction-picture the relevant operator
$\gamopi(\tau;\{\gamma\})$ can be expressed as a linear combination of
Schr\"odinger-picture operators $\gamopj$ \bea
\label{iaoperator}
\gamopi(\tau;\{\gamma\})&=& \Undag{\tau}\,\gamopi \,\Un{\tau}=
{K_{\{\gamma\}}(\tau)_i}^j\, \gamopj.  \eea Third, we define an
interaction-picture Hamiltonian by \bea
        \label{iaham}
        \hog{}(\tau)&=&
        \Undag{\tau}\,\hogSchroedinger{\tau}{}\,\Un{\tau}.  \eea With
        these definitions, one can evaluate the integrand along the
        characteristic. Details of this calculation are outlined in
        Appendix~{\ref{appendixA}}.  Thus, within the limit of weak
        interactions and the Markov approximation, one obtains
        the first two contributions to the coarse-grained density
        operator as 
\bea
\label{coarsegrainedop}
\sigam{}&=&\sign{} -i\,\int_{-\infty}^0 d\tau\, e^{\eta \tau}\, \left(
  \comut{\hog{}(\tau)}{\sign{}}+\right.\nonumber\\
&+&\left.
\trace{\comut{\hog{}(\tau)}{\gamopi}\,\sign{}} 
\,\dd{\gamma_i}\sign{}
\right)+{\cal{O}}[2].
\eea It is easy to show that within these approximations the
coarse-grained density operator is Hermitian, and that it yields
exactly the same expectation values as the reference distribution
alone, as defined in Eq.~(\ref{coord2}).

\subsection{Quantum kinetic equations}
\label{sec2d}
With this first order result for the coarse-grained density operator,
we can now address the second part of our ``boot-strapping''
procedure: the kinetic evolution of an expectation value
$\av{\hat{o}}$ subject to this coarse-grained density operator.  The
kinetic equations are obtained by averaging Heisenberg's
equation with the coarse-grained density operator, as given in
Eq.~(\ref{dynamicmean}).  As we are interested in a power series
expansion of the kinetic equations, again we decompose the total
Hamilton operator $\widehat{H}$ and the coarse-grained density
operator $\sigam{}$ into its various contributions in terms of the
interaction strength, i.e.,  \bea \frac{d}{dt}\av{\hat{o}}&=& i
\,\trace{\comut{\widehat{H}}{\hat{o}}
  \,\sigam{(t)}}= \\
&=&i \,\trace{\comut{\hng{(t)}+\hog{(t)}}{\hat{o}}\,
  \sum_{l=0}^{\infty}\sigam{(t)}^{(l)}}.\nonumber \eea By grouping the
individual terms, one finds \bea
\label{kineticpowerseries}
\frac{d}{dt}\av{\hat{o}} &=&\sum_{l=0}^{\infty}
R_{\{\gamma(t)\}}^{(l)}[\hat{o}]+ L_{\{\gamma(t)\}}^{(l+1)}[\hat{o}],
\eea where we have introduced linear Liouville operators \bea
R_{\{\gamma\}}^{(l\ge0)}[\hat{o}]&=&
i \,\trace{\comut{ \hng{}}{\hat{o}}\,\sigam{}^{(l)}},\\
L_{\{\gamma\}}^{(l>0)}[\hat{o}]&=& i \,\trace{\comut{
    \hog{}}{\hat{o}}\,\sigam{}^{(l-1)}}.  \eea

\subsubsection{General observables}
To obtain practically applicable approximations for the kinetic
equation, we will truncate these series at low order, i.e., first or
second order.  In the case of a general operator $\hat{o}$ that can
not be represented by a linear combination of relevant operators, i.e.,
$\hat{o}\notin \text{Span}(\{\gamopi\,|\,i\in {\mathcal I}\})$ this
means that \bea
\label{kineticgeneral}
\frac{d}{dt}\av{\hat{o}} &=&R_{\{\gamma(t)\}}^{(0)}[\hat{o}]+
L_{\{\gamma(t)\}}^{(1)}[\hat{o}]+\\
&+&R_{\{\gamma(t)\}}^{(1)}[\hat{o}]+
L_{\{\gamma(t)\}}^{(2)}[\hat{o}]+{\cal O}[2],\nonumber \eea which is
correct up to first order. At first glance, it seems to be
inconsistent to include also the second order contribution $L^{(2)}$
in this first order expression.  However, a closer inspection shows
that this particular kinetic equation Eq.~(\ref{kineticgeneral})
preserves all constants of motion.  In other words, if the operator
$\hat{o}$ is an exact symmetry \bea \comut{\widehat{H}}{\hat{o}}&=&0,
\eea then, according to Eq.~({\ref{kineticgeneral}}), all initial
averages are conserved \bea
\label{conservedav}
\frac{d}{dt}\av{\hat{o}}&=&0.  \eea This is particularly interesting
if we consider constants of motion related to number ($\widehat{N},
\widehat{N}^2,\ldots$) and energy
$(\widehat{H},\widehat{H}^2,\ldots)$, as implemented in many recent
BEC experiments. If the systems are prepared initially in one of the
standard thermo-dynamical ensembles: micro-canonical ($\Delta N=0$,
$\Delta E=0$), canonical ($\Delta N=0$, $\Delta E$), or
grand-canonical ($\Delta N$, $\Delta E$), then these properties are
preserved in time.

\onecolumn
\narrowtext
\subsubsection{Master variables}
In the previous section the kinetic evolution of a general operator
was examined. Now, we focus on the set of relevant operators
$\{\gamopi\}$.  The kinetic equations for the corresponding
expectation values $\{\gami\}$ constitute a self-consistent set of
master-equations that determine the system's evolution.  Again, we are
only interested in a low order truncation of
Eq.~(\ref{kineticpowerseries}).  Due to the requirement that the
reference distribution yields the same expectation values as the
unexpanded state Eqs.~(\ref{coord2},\ref{struct}), all contributions
of $R_{\{\gamma\}}^{(l>0)}[\gamopi]=0$ vanish identically.  Thus, we
find a quantum kinetic master-equation correct up to third order, i.e.,
\bea
\label{generalkineticeq}
\frac{d}{dt}\gami(t)&=& R_{\{\gamma(t)\}}^{(0)}[\gamopi]+
L_{\{\gamma(t)\}}^{(1)}[\gamopi]+
L_{\{\gamma(t)\}}^{(2)}[\gamopi]+{\cal{O}}[3].\nonumber\\
\eea By using the explicit expression for the coarse-grained density
matrix Eq.~(\ref{coarsegrainedop}), we find the following 
quantum-Boltzmann equation 
\widetext
\bea
\label{qBoltzmann}
\lefteqn{ \frac{d}{dt}\gami(t)=
  i\,\trace{\comut{\widehat{H}}{\gamopi}\, \sign{(t)}}+}\\
&&-\int_{-\infty}^{0}d\tau\, e^{\eta \tau}\, \text{Tr}\left\{
  \sign{(t)}\, \left[
    \hog{(t)}(\tau),\,\comut{\hog{(t)}(0)}{\gamopi}
    +\gamopj\, \left( i
      \dd{\gamj}L_{\{\gamma(t)\}}^{(1)}[\gamopi] +\,
      \trace{\comut{ \dd{\gamj}{\hog{(t)}(0)}}
        {\gamopi}\,\sign{(t)}} \right) \right]
\right\},\nonumber \eea 
\narrowtext
where $\hog{}(\tau)$ is the
interaction picture Hamiltonian as defined in
Eq.~(\ref{iaham}).
        
        There are two interesting features in Eq.~(\ref{qBoltzmann}):
        First, the coherent part depends only on the total Hamiltonian
        $\widehat{H}$.  Eventually, this part will determine the
        evolution of the expectation values subject to the external
        trapping-, and the mean-field potentials. Consequently, all
        derived mean-field potentials are invariant with respect to a
        particular partitioning of the total energy $\widehat{H}$, as
        introduced by the renormalization potential $\Qog{}$ (see.
        Eq.~(\ref{hn})).  Second, the kinetic equation is strictly
        local in time, i.e., Markovian.  The validity criterion
        for this result is a correlation time of the energy
        fluctuations which is much shorter than the time scale upon
        which the expectation values evolve and this is the case for a
        weakly interacting dilute gas.
\twocolumn
\section{Kinetic Equations for Mean Fields, Normal and Anomalous 
  Fluctuations}
\label{kineticequ}
In the previous sections, we determined the form of a coarse-grained
density operator as a functional of a certain set of mean values. In
turn, the temporal evolution of the mean values is determined by the
kinetic equations Eq.~(\ref{qBoltzmann}).  In this section, we apply
these general results to the particular situation of weakly
interacting, low temperature bosonic gas.  Thus, we have to determine
the predominant energy contributions to the system Hamiltonian
$\widehat{H}$, as well as, decide on a relevant set of operators
$\{\gamopi\}$.

\subsection{The dynamical evolution}
\label{sec3a}
In second quantization, the removal or addition of a particle from or
to a position $\bx$ is described by the action of quantum field
operators $\psiop{\bx}$, $\psiopd{\bx}$ on the corresponding quantum
state.  As we are considering bosonic particles, these fields have to
obey a commutation rule in order to comply with the symmetrization
postulate:

\bea \delta(\bx-\by)&=&\comut{\psiop{\bx}}{\psiopd{\by}}.  \eea

So far, we have introduced the quantum field $\psiop{}$ in a position
representation $\psiop{\bx}$. However, in performing actual
calculations other representations are often more favorable, for
example, bare-harmonic oscillator-, or self-consistent Hartree-Fock states. 
Consequently, we will delay that choice and
work with a yet unspecified basis that spans the same single-particle
Hilbert space ${\mathcal{H}}=\text{Span}(\{\ket{q_1}\,|\, q_1 \in
{\cal Q}\})$.  Here, the index set ${\cal Q}$ encompasses all possible
single-particle quantum numbers triples $q_1$.  In this generic basis,
the removal of a particle from a position $\bx$ transforms into \bea
\label{sumconv}
\psiop{\bx}&=&\sum_{q_1}\scal{\bx}{q_1} \,\aop{q_1}
\equiv \scal{\bx}{1} \,\aop{1},\\
\delta_{q_1,q_2}&=&\left[ \aop{q_1},\aopd{q_2} \right].  \eea Here,
$\aop{q_1}$ denotes a bosonic operator that removes a particle from a
general mode $\ket{q_1}$.  To simplify the notation, we drop the name
of the dummy summation variable, i.e., $\aop{q_1}\equiv\aop{1}$, as
well as the summation symbol itself.

The temporal evolution of a weakly interacting bosonic gas is governed
by a Hamilton operator that consists of a single-particle energy in
the presence of the external trapping potential, a two-particle
interaction potential, as well as higher order contributions.  Within
the dilute gas limit, we want to assume that these higher order
contributions, mediated through three-body collisions, are unlikely to
occur, and we disregard them. Thus, we assume a Hamilton operator of
the following form: \bea
\label{generic}
\widehat{H}=\hn+\ho&=&{H^{(0)}}^{12}\,\aopd{1} \aop{2}+
\phiunpr\,\aopd{1} \aopd{2}\aop{3}\aop{4}.  \eea Here, $\hn$ denotes a
single-particle Hamilton operator with matrix elements
${H^{(0)}}^{12}=\bra{1}\bp^2/(2\,m)+V_{{\text{ext}}}(\bx)\ket{2}$.  To
be specific, we assume for convenience that the external trapping
potential is an isotropic harmonic oscillator, i.e.,
$V_{\text{ext}}(\bx)= m \omega^2\,(x^2+y^2+z^2)/2$.  In most of the
present experiments, the two-body interaction potentials
$V_{\text{bin}}(\bx_1-\bx_2)$ are repulsive, of short range, and are
described by the two-particle matrix elements: \bea
\label{2bdymatelem}
\phiunpr&=&\frac{1}{2}({\mathcal{S}})
\twobra{1}{2}V_{\text{bin}}(\bx_1-\bx_2)\twoket{3}{4},\\
\phiunpr&=&\phi^{1243}=\phi^{2134}=\phi^{2143}.  \eea Only the
symmetric part of the two-particle matrix element $\phiunpr$ is
physically relevant. Therefore, we have explicitly $(\mathcal{S})$
symmetrized it.  In the low kinetic energy range that we are
interested in, s-wave scattering is the dominant two-particle
scattering event
\cite{verhaar694,wieman495,verhaar595,verhaar197,julienne597}.  Thus,
by discarding all details of the two-particle potential, we can
describe the interaction strength with a single parameter $V_0$
related to the scattering length $a_{{\text{s}}}$ by $V_0=4\pi \hbar^2
a_{{\text{s}}}/m$. This limit corresponds to a singular interaction
potential, i.e.,  $V_{\text{bin}}(\bx_1,\bx_2)=V_0\,
\delta(\bx_1-\bx_2)$.  In the case of this delta potential, one finds
for the two-body matrix elements: \bea
\label{deltamatrixel}
\phiunpr&=& \frac{V_0}{2} \int_{-\infty}^{\infty}d^3\bx
\scal{1}{\bx}\scal{2}{\bx}\scal{\bx}{3}\scal{\bx}{4}.  \eea which need
not be symmetrized, as they are symmetric already.  However,
considering the caveats that are related to the singular functional
form of the two-particle potential \cite{huang}, we will only rely on
the existence and symmetry of the two-particle matrix elements as
defined in Eq.~({\ref{2bdymatelem}).
  
  It is interesting to note that the Hamilton operator
  Eq.~(\ref{generic}) is more general than its intended use. In case
  of trapped atoms that have several internal electronic states, it is
  only necessary to combine all external and internal quantum numbers
  into the definition of a single particle state, i.e.,
  $\ket{q_1}=\ket{n_1,l_1,m_1;F_1,M_1,\ldots}$.  This implies that all
  derived results also hold true for multi-component systems, provided
  the matrix elements ${H^{(0)}}_{12}$ and $\phiunpr$ are generalized
  accordingly (for example, double condensate mean-field equations in
  Refs.~\cite{greene597,esry298})

  The renormalization potential $\Qog{}$ accounts for the mean-field
  shifted energies that affect the single-particle propagation between
  consecutive collision events.  We have seen already in the previous
  section that the mean-field potentials, which result from a first
  order calculation Eq.~(\ref{qBoltzmann}), are invariant under the
  particular choice of a energy partitioning.  Now, by anticipating
  the results of the first-order calculation, we do not just find a
  single mean-field potential, but several: i.e., one occurring in the
  equation for mean-fields, and a different one occurring in the
  equation for fluctuations.  Furthermore, due to the inclusion of the
  anomalous fluctuations (see next section), we no longer have
  particle-number conserving mean-field potentials either.  However,
  as we will use the particle-conserving part of the mean-field
  energies to determine the ``best'' single-particle basis, we will
  also choose a renormalization potential that is number conserving,
  i.e.,  \bea
\label{renormpot}
\Qog{}&=&\phiunpr \, \aopd{1}\,Q^{23}_{\{\gamma\}} \aop{4}, \eea where
$Q^{23}_{\{\gamma\}}$ are the matrix-elements (yet to be determined).
 
\subsection{The set of relevant operators}
 \label{sec3b}
 The derivation of the kinetic equations is based on the premise that
 only a few fundamental variables determine the gas's evolution on a
 coarse-grained time scale.  These quantities will serve as master
 variables and any single time observable is linked to them.  It is an
 intricate question to determine a set of relevant variables from
 general grounds up, but we are guided by the following physical
 arguments:

 In the case of kinetic temperatures well above the transition
 temperature, it is sufficient to consider only the redistribution of
 populations $f_{q_1}=\langle\aopd{q_1}\aop{q_1}\rangle$ within
 generic quantum levels $\ket{q_1}$.  The quantum average is defined
 by Eq.~(\ref{quantumav}).

 However, as temperatures are lowered, the spatial extension of a
 single-particle wave function becomes comparable to the mean
 inter-particle distance. Thus, it will be necessary to consider
 spatial coherences as well, i.e.,  $f_{q_1,q_2}=\langle\aopd{q_2}
 \aop{q_1}\rangle$. Note, this is still a single-particle quantity.

 On the other hand, the most salient and fascinating feature of
 Bose-Einstein condensation is the formation of a macroscopic
 many-particle mean-field $\al{q_1}=\langle \aop{q_1}\rangle$.  By
 now, this is an experimentally well established fact and to a large
 degree the mean-field is described by the Gross-Pitaevskii
 equation.  Moreover, there are theoretical predictions
 \cite{griffin496,stringari698} indicating that anomalous fluctuations
 play a significant role as well.  Consequently, we will also consider
 anomalous averages $m_{q_1,q_2}=\langle\aop{q_1} \aop{q_2}\rangle$ as
 independent relevant variables.
 
 Thus, guided by the forgoing arguments, we choose the set of relevant
 operators as \bea
\label{relevantop}
\{\gamopi\,|\,i \in {\cal I} \} &=&\{ {\bf 1},\,
\aop{q_1},\,\aopd{q_2},\\
\fsop{q_1 q_2}&=&(\aopd{q_2}-\ald{q_2})(\aop{q_1}-\al{q_1}),\nonumber\\
\msop{q_1 q_2}&=&
(\aop{q_1}-\al{q_1})(\aop{q_2}-\al{q_2}),\nonumber\\
\nsop{q_1 q_2}&=& (\aopd{q_1}-\ald{q_1})(\aopd{q_2}-\ald{q_2})\,| \,
q_1,q_2 \in {\cal Q}\}.\nonumber \eea and denote the corresponding
expectation values $\gami=\text{Tr}\{\gamopi \, \sigam{}\}$ by \bea
\{\gami\,|\,i \in {\cal I}\}
&=&\{1,\, \al{q_1},\,\ald{q_2},\\
&&\fs{q_1 q_2},\,\ms{q_1\,q_2},\, \ns{q_1 q_2}\,| \,q_1,q_2 \in {\cal
  Q}\}.\nonumber \eea With this set of independent variables, i.e.,
the mean fields and the fluctuations around them, we can parameterize
the reference distribution as \bea
\label{refdistexplicit}
\signal&=&{\text{exp}}\left(\Omega_{\{\fs{},\ms{},\ns{}\}}
  -\fsop{1 2}\, \Upsq{{1 2}}+\right.\nonumber\\
&&\left.-\msop{1 2}\, \Lamsq{{1 2}} -\nsop{1 2}\, \Lamsq{{1
      2}^{\,\ast}}\right).  \eea Here, we have used the implicit
summation convention described in Eq.~(\ref{sumconv}). The conjugate
thermo-dynamic coordinates $\{ \Omega, \Upsilon, \Lambda\}$ are
implicitly defined by the quantum averages \bea
\aval{\hat{o}}&=&\trace{\hat{o}\,\signal}, \eea that is \bea
\begin{array}{ll}
  \aval{{\bf 1}}=1, & \aval{\aop{1}}=\al{1},\\
  \aval{\fsop{1 2}}=\fs{1 2},& \aval{\msop{1 2}}=\ms{1 2},
\end{array}
\eea as well as their complex conjugates $\ns{1 2}=\ms{1 2}^\ast$.
The average of any other multiple operator product, occurring during
the evaluation of the kinetic equations, are greatly simplified by the
Gau{\ss}ian structure of the reference distribution.  A set of
factorization rules, known as Wick's theorem \cite{zubarev1},
can be derived and its main results are outlined in
Appendix~\ref{wick}.

\subsection{Renormalized master equations}
\label{sec3c}
In this section we present the results of applying the kinetic master
equations Eq.~(\ref{qBoltzmann}) to the set of relevant operators
defined in Eq.~(\ref{relevantop}).  Within the limits of the physical
approximations we have obtained a self-consistent set of equations for
the mean-field amplitude $\al{}$ that generalizes the Gross-Pitaevskii
 equation, a quantum Boltzmann equation for
the normal fluctuations (depletion) $\fs{}$ and the anomalous
fluctuations $\ms{}$.  The large number of individual algebraic
transformations ($\approx 10000$) that are necessary to obtain the
final result prohibits attempts to evaluate the collision terms
manually.  Therefore, we developed a symbolic algebra package that
performs the required calculations.  The presentation of the final
results of this calculation is greatly simplified by introducing the
following single-particle Hilbert-space vectors (co-, contra-variant) \bea
\begin{array}{ll}
  \av{\psiop{}} \equiv \mf{} =\al{1}\,\ket{1}, & \av{\psiop{}}^\dag
  \equiv \mfd{} =\ald{1}\,\bra{1},
\end{array}
\eea normal operators [tensor rank (1,1)]
 \bea
\begin{array}{ll}       
  \fs{}=\fs{12}\,\ket{1}\bra{2}, &
  \fc{}=\ald{2}\al{1}\,\ket{1}\bra{2},
\end{array}
\eea 
pseudo operators [tensor rank (2,0)] \bea
\begin{array}{ll}       
  \ms{}=\ms{12}\,\ket{1}\ket{2}, & \mc{}=\al{2}\al{1}\,\ket{1}\ket{2},
\end{array}
\eea and their Hermitian conjugates $\ns{}={\ms{}}^\dag$,
$\nc{}={\mc{}}^\dag$.
\subsubsection{Mean fields}
For the macroscopic many-particle mean field $\mf$, we find \bea
\label{meanfield}
\frac{d}{dt} \mf{}&=&-i\,\bigl (H^{(0)}+1\,\Uc+2\,\Usq\,\bigr ) \mf+\\
&&-i\,\Vsq\,\angle\,\mfd+\Ltwo{\aop{}},\nonumber \eea with a collision
term given by \bea
\label{l2a}
\Ltwo{\aop{}}&=&\bigl (
\Gcoll{\fs{} \fs{} \fspo{}}  -\Gcoll{\fspo{} \fspo{} \fs{}}+\\
&+&2\, \Gcoll{\fs{} \ms{} \ns{}} -2\, \Gcoll{\fspo{} \ms{} \ns{}}
\bigr)\,\mf{}+\nonumber\\
&+&2\,\bigl( \Gcoll{\fs{} \ms{} \fspo{}}-\Gcoll{\fspo{} \ms{} \fs{}}
\bigr)\,\angle\,\mfd.\nonumber \eea First of all, the field evolves
unitarily in the nonlinear hermitian Hamilton operator which consists
of a free part $H^{(0)}$, determined by the external trapping
potential and any other applied electro-magnetic fields. The second and third 
contribution are the collision induced mean field potentials, denoted
by $\Uc$ and $\Usq$.  While the first of these potentials is proportional the
mean field density itself, the second potential $\Usq$ reflects the
influence of the normal fluctuation upon the mean field.  
It is
important to note the different weighting factors $1$ and $2$
multiplying the potentials.  They arise from the different quantum
statistical fluctuation properties of a c-number mean field and a
normal single particle density.  Exactly the same weighting factors
are also found with the variational Hartree-Fock-Bogoliubov
approach \cite{griffin496}.  The mean field potential is defined in
terms of the two-particle interaction matrix elements
Eq.~(\ref{2bdymatelem}) and a single-particle density operator $\ft{}$
that  can be either $\fc{}$ or $\fs{}$ 
\bea U_{\ft{}}&=& 2\,\phipr
\,\ft{3^\prime 2^\prime}\,\ket{1}\bra{4^\prime}.  \eea 
Due to the
Hermiticity of the two-particle interaction energy and the positivity
of the single particle density, it is also self-adjoint
\footnote{This  potential $U_{\ft{}}$ is not related to the single-particle 
propagator $\Un{\tau}$ defined in Eq.~(\ref{propagator})}, i.e.,
$U_{\ft{}}=U_{\ft{}}^\dag$.  It is interesting to see that in the case
of a delta-potential Eq.~(\ref{deltamatrixel}) and a scalar field
(i.e., no internal degrees of freedom), the mean field potential
reduces to the well known potential energy density that is
proportional to the local mean field density \bea
\bra{\bx}\Uc\ket{\by}&=&V_0 \,\delta(\bx -\by)
\left|\scal{\bx}{\al{}}\right|^2.  \eea The fourth  linear collisional
contribution is proportional to the anomalous coupling strength
$\Vsq$ contracted ($\angle$) with the adjoint field $\mfd$.  In
general, we obtain the contraction of two tensor fields $A \angle B$,
from a basis representation of the two fields and a subsequent
contraction of the last index of $A$ with the last index of $B$.

This non-Hermitian coupling is in general mediated by an anomalous
average $\mt{}$, and explicitly given by \bea V_{\mt{}}&=& 2\,\phipr
\,\mt{3^\prime 4^\prime}\,\ket{1}\ket{2^\prime}.  \eea In here,
$\mt{}$ stands for any anomalous average, either $\mc{}$ or $\ms{}$.
From the definition of the anomalous coupling, it can be seen easily
that $V_{\mt{}}=V_{\mt{}}^\top$ is symmetric.

All of the remaining terms in Eq.~(\ref{l2a}) are second order
collisional contributions.  They always appear in pairs where one term
corresponds to an in-process while the sign reversed companion
describes a loss out of the field.  A closer inspection reveals that
there are essentially four types of processes occurring, i.e., collision
events that involve zero to three anomalous averages.  Furthermore,
one finds that a normal fluctuation $\fs{}$ on the in-side is always
accompanied by a bosonically enhanced $\fspo{}$ on the out-side, and
vice versa. On the other hand, whenever a mean field density $\fc{}$,
an anomalous mean field density $\mc{}$, or an anomalous fluctuation
$\ms{}$ occurs in an in-process they appear unaltered on the
out-process.  This behavior is analogous to atomic transition rates
described by the Einstein A- and B coefficients which can be
attributed to stimulated absorption- and emission, as well as
spontaneous emission processes.  The fact that the mean-field is never
bosonically enhanced supports the interpretation that the mean-field
acts as a classical driving field.

In detail, these collisions operators are described by the following
operators and pseudo-operators \bea
\begin{array}{l}
  \Gcoll{\ft{} \ft{} \ft{}}= 8\,\phipr \phippr{\eta} \ft{3^\prime
    1^\ppr} \ft{4^\prime 2^\ppr}
  \ft{4^\ppr 2^\prime}\,\ket{1} \bra{3^\ppr},\\
  \Gcoll{\ft{} \mt{} \ft{}}= 8\,\phipr \phippr{\eta} \ft{3^\prime
    1^\ppr} \mt{4^\prime 3^\ppr}
  \ft{4^\ppr 2^\prime }\,\ket{1} \ket{2^\ppr},\\
  \Gcoll{\ft{} \mt{} \nt{}}= 8\,\phipr \phippr{\eta} \ft{3^\prime
    1^\ppr} \mt{4^\prime 3^\ppr}
  \nt{2^\ppr 2^\prime}\,\ket{1} \bra{4^\ppr},\\
  \Gcoll{\mt{} \mt{} \nt{}}= 8\,\phipr \phippr{\eta} \mt{3^\prime
    4^\ppr} \mt{4^\prime 3^\ppr} \nt{2^\ppr 2^\prime}\,\ket{1}
  \ket{1^\ppr}.
\end{array}\nonumber\\
\eea From the time average over the interaction picture Hamilton
operator that appears in the kinetic equation Eq.~(\ref{qBoltzmann}),
one obtains an approximately energy conserving two-particle matrix
element.  \bea
\label{phieta}
\lefteqn{ \phippr{\eta}(t)=\int_{-\infty}^{0} d\tau \,e^{\eta\tau}\,
  \phi^{1 2 3 4}\, \times} \\
&&{{K_{\{\gamma (t)\}}^\dag}(\tau)^{1^\ppr}}_1
{{K_{\{\gamma(t)\}}^\dag}(\tau)^{2^\ppr}}_2
{{K_{\{\gamma(t)\}}(\tau)}^{3^\ppr}}_3
{{K_{\{\gamma(t)\}}}(\tau)^{4^\ppr}}_4.\nonumber \eea The restricted
propagator that has been used here
is explicitly given by the time-ordered exponential \bea
K_{\{\gamma(t)\}}(\tau)&=&T \,e^{i\int_{\tau}^{0} ds \,( H^{(0)}+Q_{
    \{\overline{\gamma}(s;\{\gamma(t)\})\}})}.  \eea where we obtained
the operators $H^{(0)}$ and $Q_{\{\gamma\}}$ from the matrix elements
given in Eqs.~(\ref{generic},\ref{renormpot}).

To see qualitatively why $\phippr{\eta}$ is essentially non-zero only
on the energy-shell of thickness $\eta$, it is useful to represent the
restricted propagator with respect to the eigen-states of \bea
(H^{(0)}+Q_{ \{\gamma(t)\}})\ket{1}&=&\varepsilon_{1}(t)\ket{1}.  \eea
By assuming that the energy levels change adiabatically slow, one
obtains approximately \bea \phippr{\eta}&=&\phippr{}\, (\pi
\delta_{\eta}(\Delta_{1^\ppr 2^\ppr 3^\ppr 4^\ppr}) +i
{\mathcal{P}_\eta}\frac{1}{\Delta_{1^\ppr 2^\ppr 3^\ppr 4^\ppr}}),
\nonumber\\
\eea which is non-zero only if the energy difference $\Delta_{1^\ppr
  2^\ppr 3^\ppr 4^\ppr}=\varepsilon_{1^\ppr}(t)+
\varepsilon_{2^\ppr}(t)-\varepsilon_{3^\ppr}(t)-\varepsilon_{4^\ppr}(t)$
is smaller than $\eta$.  
\bea 
\lim_{\eta\rightarrow 0_+}\frac{1}{\eta-i \Delta}&=& 
\pi\,\delta_\eta(\Delta)+i\,{\mathcal{P}}_\eta\frac{1}{\Delta}, 
\eea
This result is analogous to the second order Born-Markov
approximation \cite{gardiner}.
\onecolumn 
\narrowtext
\subsubsection{Normal fluctuations}
The kinetic equation for the normal fluctuations (depletion)
generalizes the quantum-Boltzmann equation found in many
textbooks \cite{peletminskii,zubarev1} \bea
\label{normalfluct}
\lefteqn{ \frac{d}{dt}\fs{}= -i\,\left[ H^{(0)}+2\,\Uc+2\,\Usq,
    \fs{}\right]+}\\
&&\quad-i\,\Vm\,\angle\, \ns{}+i\,\ms{}\,\angle\, \Vm^\dag+
\Ltwo{\fsop{}}, \nonumber \eea with second order collisional
contributions \widetext \bea
\label{l2ada}
\lefteqn{\Ltwo{\fsop{}}=\Bigl ( \Gcoll{\fs{} \fs{} \fspo{}}+
  2\,\Gcoll{\fc{} \fs{} \fspo{}} +\Gcoll{\fs{} \fs{} \fc{}}
  +2\,\Gcoll{\fs{} (\mc{}+\ms{}) \ns{}}+ 2\,\Gcoll{\fs{} \ms{} \nc{}}
  +2\,\Gcoll{\fc{} \ms{} \ns{}}
  \Bigr)\fspo{}+}\\
&&\quad-\Bigl( \Gcoll{\fspo{} \fspo{} \fs{}}+ 2\,\Gcoll{\fc{} \fspo{}
  \fs{}} +\Gcoll{\fspo{} \fspo{} \fc{}}+ 2\,\Gcoll{\fspo{}
  (\mc{}+\ms{}) \ns{}}+ 2\,\Gcoll{\fspo{} \ms{} \nc{}}
+2\,\Gcoll{\fc{} \ms{} \ns{}}
\Bigr)\fs{}+\nonumber\\
&&\quad+2\Bigl ( \Gcoll{\fs{} (\mc{}+\ms{}) \fspo{}}+ \Gcoll{\fc{}
  \ms{} \fspo{}} +\Gcoll{\fs{} \ms{} \fc{}} -\Gcoll{\fspo{}
  (\mc{}+\ms{}) \fs{}}- \Gcoll{\fc{} \ms{} \fs{}} -\Gcoll{\fspo{}
  \ms{} \fc{}} \Bigr)\,\angle\,\ns{}+{\text{h.c.}}\nonumber \eea
\narrowtext 
First, one finds a unitary evolution in the presence of
the external trapping-, the mean-field-, and the normal potential.
Both of the self-induced potentials, $\Uc$ and $\Usq$ are weighted by
a common factor of $2$ (compare \cite{griffin496}).  
This is in contrast to the weighting factors
appearing the mean-field Hamilton operator Eq.~(\ref{meanfield}).  But
again, this fact can be traced back to different quantum statistical
properties of the mean-field and the fluctuations.  Second, it can be
seen that the anomalous coupling strength is now proportional to the
total anomalous average, i.e., $\mc{}+\ms{}$.  Third, in the absence of
any mean-fields or anomalous averages the second order contribution in
Eq~(\ref{l2ada}) reduces to the well known Boltzmann collision
term \bea \Gcoll{\fs{} \fs{} \fspo{}}\fspo{}- \Gcoll{\fspo{} \fspo{}
  \fs{}}\fs{}.  \eea By further assuming that the normal fluctuations
are predominantly diagonal in an energy eigen-basis defined by the
non-linear Hamilton operator of Eq.~(\ref{normalfluct}) (ergodic
hypothesis), one recovers the Bose-Einstein distribution as the
stationary distribution of particles within the quantum levels.
However, the presence of the mean-field, as well as the anomalous
averages lead to additional collision processes that must not be
ignored in general.  Eventually, these processes will lead to a
self-consistent equilibrium partition of particles between
mean-fields, normal- and anomalous fluctuations. While a detailed
numerical self-consistent solution of the set of kinetic equations is
still under investigation, it is important to see that the total
particle number $\av{\hat{N}}={\text Tr}\{\fc{}\}+{\text Tr}\{\fs{}\}$
is always conserved [compare Eq.~(\ref{conservedav})] \bea
\frac{d}{dt}\av{\hat{N}}=0.  \eea
\subsubsection{Anomalous fluctuations}
In contrast to the normal fluctuations, the anomalous fluctuations do
not evolve unitarily but rather as a tensor of rank (2,0). Both, left
and right generators of the time-evolution are identical to the
Hamilton operator of the normal fluctuations.  \widetext \bea
\label{anomalfluct}
\frac{d}{dt}\ms{}&=& -i\,\bigl (H^{(0)}+2\,\Uc+2\,\Usq \bigr
)\,\angle\, \ms{}
-i \,\ms{}\,\angle\, \bigl ( H^{(0)}+2\,\Uc+2\,\Usq \bigr)+\\
&&-i\,\Vm\,\angle \, \fspo{}-i\,\fs{}\,\angle\,
\Vm+\Ltwo{\msop{}},\nonumber \eea \bea \lefteqn{ \Ltwo{\msop{}}= \Bigl
  ( \Gcoll{\fs{} \fs{} (1+\fs{})}+\Gcoll{\fs{} \fs{} \fc{}}+
  2\,\Gcoll{\fs{} \fc{} \fspo{}} +2\,\Gcoll{\fs{} \ms{} (\nc{}+\ns{})}
  +2\,\Gcoll{\fs{} \mc{} \ns{}}
  \Bigr)\,\angle\,\ms{}+}\\
&& \quad-\Bigl ( \Gcoll{\fspo{} \fspo{} \fs{}}+ \Gcoll{\fspo{} \fspo{}
  \fc{}}+2\,\Gcoll{\fspo{} \fc{} \fs{}} +2\,\Gcoll{\fspo{} \ms{}
  (\nc{}+\ns{})} +2\,\Gcoll{\fspo{} \mc{} \ns{}}
\Bigr)\,\angle\,\ms{}+\nonumber\\
&&\quad+ \Bigl( 2\,\Gcoll{\fs{} (\mc{}+\ms{}) \fspo{}}+
2\,\Gcoll{\fs{} \ms{} \fc{}} +2\, \Gcoll{\fc{} \ms{} \fspo{}}
+\Gcoll{\ms{} \ms{} (\nc{}+\ns{})}+2\, \Gcoll{\ms{} \mc{} \ns{}}
\Bigr)\,\angle\,\fs{}+\nonumber\\
&&\quad- \Bigl( 2\,\Gcoll{\fspo{} (\mc{}+\ms{}) \fs{}}+
2\,\Gcoll{\fspo{} \ms{} \fc{}} +2\, \Gcoll{\fc{} \ms{}
  \fs{}}+\Gcoll{\ms{} \ms{} (\nc{}+\ns{})} +2\, \Gcoll{\ms{} \mc{}
  \ns{}} \Bigr)\,\angle\,\fspo{}+{\text{transp.}}\nonumber \eea
\narrowtext 

These three sets of master equations for the mean-field,
the normal-, anomalous fluctuations constitute the main result of this
article.  They unify and generalize simpler equations 
which have been obtained previously also by other methods.  
However, so far, we have  not discussed the physical
implications that will arise from a self-consistent solution of these
equations. Specifically, we need to determine the following problems:  
(I) the equilibrium distribution of particles partitioned
between mean-field, normal-, and anomalous fluctuations; 
(II) the importance and quantitative size of anomalous fluctuations; 
(III) the collisional damping rates and second order energy-shifts; 
(IV) the response of the equilibrium system to weak external perturbations, 
i.e. the collective excitation frequencies via linear response theory;
(V) critical phenomena occurring around the onset of condensation;
or, for example, 
(VI) the dynamics of the growth of the condensed phase. 
\twocolumn
\section{Outlook}
\label{sec4}
In the previous section, we have enumerated 
several quantities that need to be determined and interesting paths 
along which detailed calculations could proceed. 
We believe that amongst these issues, it will be most crucial
to address problems  (I) and (II) around $T\approx0$, first. 
On one hand, present-day experiments have established that 
the mean-field  description yields good agreement.
On the other hand,  there are various approaches to the
self-consistent equilibrium for normal and anomalous fluctuations
and not all implications have been elucidated. 

The standard route to investigate this problem is based on 
finite temperature calculations in the Hartree-Fock-Bogoliubov description.
Various schemes employing, for example, the quasi-static Popov 
approximation, or more dynamical methods that go  beyond  it 
(i.e.~the collisionless regime) are being investigated by several 
research groups.

This present, non-equilibrium approach provides an alternate route to
the stationary solution. In particular, we expect that the presence of 
a large condensed phase will lead to a strong correlation of the low energy 
part of the normal and anomalous fluctuations 
($\approx1-2$ times the chemical potential $\mu$ of the condensate), 
while the high energy tail will be mostly in detailed balance at some 
temperature $T$. 
However, such a macroscopic ``polarization'' of the low energy part
of the fluctuations can not be described within a simple ergodic hypothesis, 
and therefore requires a full quantum treatment. 

The main obstacle to overcome in numerically answering this problem
is the unfavorable scaling law of the collision operators.
From a simple operations count, one finds that there are $N^8$ summations
involved if $N$ is the number of energy levels being considered.

This burden can be alleviated by being more specific, i.e. by postulating 
a completely isotropic situation for a single condensed phase,
an isotropic trapping potential, a rotationally invariant 
initial condition, as well as a short-range central 
two-particle interaction. 
Within this simplified model, one can then decompose all involved
operators in terms of angular momentum sub-manifolds 
(i.e. irreducible tensor sets and  use of Wiger-Eckart theorem).
This assumption makes the quantum mechanical treatment of the
low energy region ($\mu\le \varepsilon\le 2\,\mu$, $N\approx 10-20$) 
feasible and will lead to a self-consistent equilibrium. 
A detailed numerical investigation is in progress 
and results will be reported.

\section*{Conclusions}
In this article, we have revisited the Chapman-Enskog-Bogoliubov
procedure of non-equilibrium statistical mechanics to describe the
kinetic evolution of a condensed bosonic gas of atoms towards
equilibrium.  Within a second order Born-Markov approximation,
we consider the collision dynamics of macroscopic mean-fields, normal
fluctuations, and anomalous averages.  In particular, we have obtained
a coupled set of master equations for these quantities that encompass
the Gross-Pitaevskii mean-field equation, as well as the
quantum-Boltzmann equation for the normal fluctuations as
limiting cases.  The mean-field potentials that are obtained from a
first order calculation are in agreement with the results of a
variational Hartree-Fock-Bogoliubov calculation.  Beyond these
first order energy shifts, we obtain second order collisional energy
shifts and damping rates that are bosonically enhanced.  We expect our
results to be valid when strong collisions are well separated in time
and when the mean-field induced energy shifts may be neglected during
a strong collision event [see Eqs.~(\ref{Dterm}), (\ref{Sterm})].
\section*{Acknowledgments}
We gratefully acknowledge stimulating discussions with P.~Zoller, 
C.~W.~Gardiner, K.~Burnett and M.~L.~Chiofalo.
\appendix
\section{}
\label{appendixA}
With the definitions for the interaction-picture representation, we
can rewrite the commutator term of Eq.~(\ref{firstorder}) as \bea
&&\Undag{\tau}\,
\comut{\hogSchroedinger{\tau}{}}{\signSchroedinger{\tau}{}}\,
\Un{\tau}=\nonumber\\
&&=\comut{\hog{}(\tau)}{\sign{}}.  \eea The trace term evaluated along
the trajectory simplifies to \bea \lefteqn{
  \trace{\comut{\hogSchroedinger{\tau}{}}{\gamopi}
    \,\signSchroedinger{\tau}{}}=}\\
&=&{K_{\{\gamma\}}(\tau)_i}^j \,
\trace{\comut{\hog{}(\tau)}{\gamopj}\,\sign{} }.\nonumber \eea
Finally, the self-tuning term of the reference distribution gives \bea
\lefteqn{ \Undag{\tau}\, \dd{\gami}
  {\sign{}}_{|\overline{\gamma}(\tau;\{\gamma\})}\,
  \Un{\tau}=}\\
&=&\left( \frac{\partial \overline{\gamma}_i (\tau;\{\gamma\})}{
    \partial \gamma_l} \right)^{-1} \left( \comut{
    \widehat{D}_{\{\gamma\}}(\tau)_{l} }{\sign{}}+
  \dd{\gamma_l}\sign{} \right),\nonumber \eea where we introduce
auxiliary operator valued vectors $\widehat{D}_{\{\gamma\}}(\tau)$ and
matrix-valued coefficients, $S_{\{\gamma\}}(\tau)$ by \bea
\widehat{D}_{\{\gamma\}}(\tau)_{i}&=&\Undag{\tau}\,
\partial_{\gamma_i} \Un{\tau},\\
{S_{\{\gamma\}}(\tau)}^{i j}&=& \left( \frac{\partial
    \overline{\gamma}_l (\tau;\{\gamma\})}{ \partial \gamma_i}
\right)^{-1} {K_{\{\gamma\}}(\tau)_l}^j.  \eea If these results are
put together, one finds for the first order correction of the
coarse-grained statistical operator \bea \sigo{}&=&
-i\,\int_{-\infty}^0 d\tau\, e^{\eta \tau}\, \left(
  \comut{\hog{}(\tau)}{\sign{}}+\right.\nonumber\\
&+& \left( \comut{\widehat{D}_{\{\gamma\}}(\tau)_i}{\sign{}}+
  \dd{\gamma_i}\sign{} \right)
S_{\{\gamma\}}(\tau)^{i j}\nonumber\\
&&\left.  \trace{ \comut{\hog{}(\tau)}{\gamopj}\, \sign{}} \right).
\eea This expression is formally equivalent to Eq.~(\ref{firstorder}).
However, a closer inspection of the $S$ and $\widehat{D}$ terms shows
that they contain higher order energy corrections induced by
renormalization energy $\widehat{Q}_{\{\gamma\}}^{(1)}$.  In
particular, from a short time Taylor-expansion one finds that \bea
\label{Dterm}
\widehat{D}_{\{\gamma\}}(\tau)&=&0 -i\,{\cal O}\left[
  \tau\,\,\partial_{\gamma_i}\widehat{Q}_{\{\gamma\}}^{(1)}\right],\\
\label{Sterm}
S_{\{\gamma\}}(\tau)^{i j}&=& \delta^{i
  j}-i\,{\cal{O}}\left[\tau\,\,\gamma_l\, \partial_{\gamma_j}
  {\astruc{\gamma}_{i}}^l\right].  \eea Consequently, we will
disregard the effect of the mean-field onto the temporal evolution of
$\widehat{D}_{\{\gamma\}}(\tau)$ and $S_{\{\gamma\}}(\tau)$ during a
strong collision event and replace them by their ``bare'' values
attained in the absence of the mean-field shift.

\section{A generalized Wick's theorem}
\label{wick}
The Gau{\ss}ian structure of the reference distribution
Eq.~(\ref{refdistexplicit}) is particularly useful, as it permits the
systematic application of Wick's theorem \cite{zubarev1}.  This
is a set of rules to efficiently evaluate quantum averages for
multiple operator products as \bea
\label{multop}
\langle \kpsiop{1} \kpsiop{2} \ldots \kpsiop{l}
\rangle_{\{\al{},\ald{},\fs{},\ms{},\ns{}\}}.  \eea In this average,
for example, the operator $\kpsiop{1}$ represents either an operator
$\aop{1}$ or $\aopd{1}$.

First, the displacement rule shifts any operator $\kpsiop{1}$ by its
c-number expectation value $\psi_{1}$ which is either $\al{1}$ or
$\ald{1}$, and replaces the quantum average by an average that has
zero mean values: \bea \lefteqn{ \aval{ \kpsiop{1} \kpsiop{2}\ldots
    \kpsiop{l}}=}\\
&=&\avnull{(\kpsiop{1}+\psi_{1}) (\kpsiop{2}+\psi_{2})\ldots
  (\kpsiop{n}+\psi_{l})}.\nonumber \eea Second, after expanding the
multiple products, one can discard all averages that involve an odd
numbers of operators: \bea \avnull{ \kpsiop{1} \kpsiop{2}\ldots
  \kpsiop{2s+1}}&=&0.  \eea And third, for the remaining averages, one
can use the Gau{\ss}ian factorization rule: \bea \lefteqn{ \avnull{
    \kpsiop{1} \kpsiop{2}\ldots
    \kpsiop{2s}}=}\\
&=&\avnull{ \kpsiop{1} \kpsiop{2}} \avnull{ \kpsiop{3}\ldots
  \kpsiop{2s}}+\nonumber\\
&+&\avnull{ \kpsiop{1} \kpsiop{3}} \avnull{ \kpsiop{2} \kpsiop{4}
  \ldots
  \kpsiop{2s}}+\nonumber\\
&&\quad\quad\vdots\nonumber\\
&+&\avnull{ \kpsiop{1} \kpsiop{2s}} \avnull{ \kpsiop{2} \ldots
  \kpsiop{2s-1}}.\nonumber \eea By proceeding recursively, one has
finally evaluated the complete multiple operator average
Eq.~(\ref{multop}).
                               
\hyphenation{Post-Script Sprin-ger}


\end{document}